\documentclass[aip,twocolumn,letterpaper]{revtex4-2}
\usepackage[margin=0.75in]{geometry}
\usepackage{graphicx}
\usepackage{amsmath}
\usepackage{amsfonts}
\usepackage{amssymb}
\usepackage{xcolor}
\usepackage{hyperref}
\usepackage{appendix}
\usepackage{booktabs}
\usepackage{multirow}

\newcommand{\Poincare}{Poincar\'{e}\;}
\selectfont

\begin{document}
\title{Stellarator divertor design by optimizing coils for surfaces with sharp corners}
\author{Todd Elder\textsuperscript{1}, Matt Landreman\textsuperscript{1}, Chris Smiet \textsuperscript{2}, Bob Davies \textsuperscript{3}}
\affiliation{University of Maryland, College Park, MD, 20737}
\affiliation{École Polytechnique Fédérale de Lausanne, Lausanne, Switzerland}
\affiliation{Max Planck Institute for Plasma Physics, Greifswald, Germany}
\affiliation{\textsuperscript{}Corresponding author: Todd Elder, \textit{telder1@umd.edu}}
\begin{abstract}

In stellarators, achieving effective divertor configurations is challenging due to the three-dimensional nature of the magnetic fields, which often leads to chaotic field lines and fuzzy separatrices. This work presents a novel approach to directly optimize modular stellarator coils for a sharp X-point divertor topology akin to the Large Helical Device's (LHD) helical divertor using a target plasma surface with sharp corners. By minimizing the normal magnetic field component on this surface, we target a clean separatrix with minimal chaos. 
Notably, this approach demonstrates the first LHD-like helical divertor design using optimized modular coils instead of helical coils.
Separatrices are produced with significantly lower chaos than in LHD, demonstrating that a wide chaotic layer is not intrinsic to the helical divertor.
Additional optimization methods are implemented to improve engineering feasibility of the coils and reduce chaos, including weighted quadrature and manifold optimization, a method which does not rely on normal field minimization. The results outline several new strategies for divertor design in stellarators, though it remains to achieve these edge divertor features at the same time as internal field qualities like quasisymmetry.
\end{abstract}
\date{\today}
\maketitle

\section{Introduction \label{sec:Intro}}

Divertors play a critical role in magnetic confinement fusion devices by managing plasma exhaust, including helium ash from deuterium-tritium reactions, impurities, and neutrals. The divertor serves as a barrier between the hot core plasma (temperatures around $\sim$10 keV) and the device walls, reducing temperatures to $\sim$1-2 eV at the divertor targets. Effective divertors must handle high heat fluxes, localize neutrals so they can be efficiently pumped, and maintain plasma purity by preventing impurities from the wall from reaching the core \cite{burnett1958divertor,Krasheninnikov2017}.

In stellarators, non-axisymmetric magnetic fields offer flexibility but complicate divertor design. Unlike tokamaks, where axisymmetric poloidal X-point divertors are standard, stellarator divertors often rely on resonant islands, ergodic layers, or non-resonant structures, each with trade-offs in sensitivity to plasma currents and chaos levels \cite{feng2011comparison}. Three primary types have been considered: island divertors\cite{karger1977resonant} (e.g., Wendelstein 7-X\cite{renner2002divertor}), helical divertors (e.g., the Large Helical Device LHD \cite{morisaki2006review}) , and non-resonant divertors\cite{punjabi2020simulation} (e.g., the Helically Symmetric eXperiment HSX \cite{bader2017hsx} and the Compact Toroidal Hybrid \cite{garcia2023exploration}) . Desirable properties include resilience to plasma-current-induced field changes and the ability to achieve high-recycling or detached regimes.

This work introduces a coil optimization method to achieve a helical divertor topology resembling the LHD divertor, but with modular coils, and with reduced stochasticity. 
The key idea is to perform standard coil optimization but using a target plasma shape with toroidally continuous sharp edges, which become the X-lines of a separatrix.
Further efforts to refine the initial result for improved divertor properties and reduced coil complexity are shown, illustrating effective strategies for stellarator divertor design.

The helical divertor offers the advantage of tight baffling, enabling high pumping efficiency and effective neutral and impurity removal \cite{morisaki2013initial,motojima2017establishment}. 
Historically, it has been assumed that such helical divertor topologies require continuous helical coils, which are considered challenging for assembly and less reactor-relevant than modular coils until superconducting joint technology advances significantly. 
For instance, modular coils could be built in a factory and transported to a reactor site, whereas helical coils (without joints) must be wound directly on the vacuum vessel.
Moreover, LHD's divertor exhibits a highly stochastic separatrix due to high shear and overlapping resonances, which is thought to cause momentum loss and hinder detachment \cite{feng2011comparison}. In this work, we demonstrate that an LHD-like helical divertor can be achieved using modular coils, with substantially reduced chaos compared to LHD, showing that a tunable range of stochasticity is possible, potentially balancing heat flux spreading with improved detachment access \cite{feng2011comparison}.

The fact that modular coils can generate flux surfaces with sharp edges that may be used for a divertor was pointed out several decades ago by Derr and Shohet \cite{derr1981modular}.
However in that work, the modular coil shapes were not optimized in the manner of modern stellarator coil design.
In more recent related work, Gaur et al. \cite{gaur2025omnigenous} optimized for plasma boundary shapes with high-curvature edges, referred to there as umbilic stellarators, with emphasis on omnigentiy and magnetohydrodynamic stability. Here, we focus on edge field structure rather than core physics properties, using a sharper-edged target surface, yielding distinct edge topologies with pronounced X-points and reduced stochasticity.

In the following section, the optimization methods are presented in detail.
Three methods are shown: a method similar to standard filament coil optimization, a weighted version of it, and a new approach termed manifold optimization.
Results are presented and compared in section \ref{sec:Results}.
We discuss and conclude in section \ref{sec:discussion}.

\section{Method \label{sec:Method}}

In the first approach, coils are optimized to produce a desired magnetic topology by minimizing the squared flux objective on a target surface with sharp corners.
Other than the shape of the target surface, the coil optimization method is standard. The squared flux objective, including a normalization by the local field magnitude, is $\int_S (B_n / B)^2 \, dA, $ where $S$ is the target surface, and $B_n = \mathbf{B} \cdot \mathbf{n}$ is the normal component of the magnetic field. 
In this study we consider only vacuum fields for simplicity, so there is no contribution from plasma currents to $\mathbf{B}$.
Terms to control the coils' length, coil-coil distance, coil-plasma distance, maximum curvature, mean squared curvature, and Gauss linking number were also added to the objective function.
The total objective function used was
\begin{eqnarray}
    &&J = \int_{S} dA\frac{\left(\mathbf{B}\cdot\mathbf{n}\right)^2}{B^2} + w_L\left(L_* - \sum_{j=1}^N L_j\right)^2 \\
&&+ w_{cc}\sum_{j=1}^N \sum_{k=1}^{j-1} \int_j d\ell \int_k d\ell \; \max\left(0, \; d_{cc*} - |\mathbf{r}_j - \mathbf{r}_k|\right)^2
\nonumber \\
&&+ w_{cs}\sum_{j=1}^N \int_j d\ell \int_{S} dA \;\max\left(0, \; d_{cs*} - |\mathbf{r}_j - \mathbf{r}_S|\right)^2 \nonumber \\
&&+ \frac{w_{\kappa}}{2} \sum_{j=1}^N \int_j
d\ell \; \max\left( 0, \; \kappa - \kappa_*\right)^2 \nonumber \\
&&+ w_{msc} \sum_{j=1}^N \max\left( 0, -c_* +\frac{1}{L_j}\int_j
d\ell \;\kappa^2 \right)^2 \nonumber \\
&&+ \sum_{j,k\ne j}\left( 
\frac{1}{4\pi}\iint \frac{\mathbf{r}_j - \mathbf{r}_k}{|\mathbf{r}_j - \mathbf{r}_k|^3}
\cdot d\mathbf{r}_j\times d\mathbf{r}_k
\right)^2. \nonumber
    \label{eqn:Coil_eng_penalty_function}
\end{eqnarray}
Here, $\ell$ is arclength along a coil, $\kappa$ is the curvature, $j$ and $k$ are indices which range over the coils, $L_j$ is the length of coil $j$, and $\mathbf{r}_j$ and $\mathbf{r}_S$ are the position vectors on coil $j$ and the target surface respectively.
The quantities $w_L$, $w_{cc}$, $w_{cs}$, $w_{\kappa}$, and $w_{msc}$ are weights.
The quantities $L_{*}$, $d_{cc*}$, $d_{cs*}$, $\kappa_*$, and $c_*$ are target values.

The target surface $S$ used in this work resembles a rotating ellipse but with sharp edges, here referred to as a `lemon' or `rotating lemon'.
It is formed by rotating a lemon-shaped cross-section (two semicircles intersecting at sharp points) toroidally. 
This target surface is inspired by LHD's helical divertor, which features two X-lines but suffers from broken separatrices due to high shear and overlapping resonances. The optimization with the rotating lemon aims to produce similar structures, but with less chaos and produced by modular coils.

A randomized search over weights and thresholds in the coil objective function was used to identify promising coil sets. Optimizations were performed using the SIMSOPT software \cite{SIMSOPT}. The randomization increases the probability of finding global optima, and provides an ensemble of solutions along the trade-off surface between the multiple objectives. 
From this ensemble we identify modular coil sets that best produce the desired X-point divertor topology.

\subsection{Target surface with sharp corners: the rotating lemon \label{sec:Method-Geometry}}

The 3D rotating lemon surface may be seen in Fig. \ref{fig:3d_wsf_coils}. A cross-section at $\phi=0.25\pi$ is shown in Fig. \ref{fig:maniopt_method}. The surface is constructed as follows.

Assuming the plasma volume is centered on a circle of major radius $R_0$, the center of each arc segment of the surface is given by the position vector
\begin{eqnarray}
\mathbf{r}_{0j} &=& \left[ R_0 + d \cos\left( \frac{n_{fp} \phi}{2} + \alpha_j \right) \right] \hat{\mathbf{e}}_R \notag\\
&& + d \sin\left( \frac{n_{fp} \phi}{2} + \alpha_j \right) \hat{\mathbf{e}}_z,
\end{eqnarray}
where $j = 1, 2$ indexes the two halves of the separatrix, with $\alpha_1 = 0$ and $\alpha_2 = \pi$. The parameter $d$ determines the separation between the centers of the two arc surfaces, $n_{fp}$ is the number of field periods, and $\hat{\mathbf{e}}_R$ and $\hat{\mathbf{e}}_z$ are cylindrical unit vectors. 

The position vector on the surface of a semicircle of radius $a$ is then
\begin{widetext}
\begin{eqnarray}
\mathbf{r}_j &=& \left[ R_0 + d \cos\left( \frac{n_{fp} \phi}{2} + \alpha_j \right) - a \cos\left( \theta + \frac{n_{fp} \phi}{2} + \alpha_j \right) \right] \hat{\mathbf{e}}_R \notag\\
&&+ \left[ d \sin\left( \frac{n_{fp} \phi}{2} + \alpha_j \right) - a \sin\left( \theta + \frac{n_{fp} \phi}{2} + \alpha_j \right) \right] \hat{\mathbf{e}}_z.
\end{eqnarray}
\end{widetext}
This parameterization enables the computation of the normal magnetic field on the surface, forming the basis for the squared flux minimization objective in coil optimization.

The coordinate $\theta$ ranges over $[-\theta_0,\theta_0]$, defining the poloidal extent of the semicircles, and controlling the length by which the surfaces extend beyond their curve of intersection. 
For the parameters used here, the two halves of the surface extend a small amount beyond the intersection.
This feature is likely not required; any surface with sharp corners can in principle be used to target the desired topology. 
Although not shown here, scans revealed that larger or smaller values of $\theta_0$ (i.e. surfaces extending farther beyond the intersection, or not extending far enough to intersect) resulting in a more chaotic separatrix produced by optimized coils. The parameters $d$ and $a$ control the angle of the surface intersection, though the dependence on these parameters was not studied here.

A single lemon configuration is used throughout this work. The parameters used are: $R_0=1$ m, $d=0.2$ m, $a=0.3$ m, $\theta_0=0.88$, $\alpha_0=0$, $\alpha_1=\pi$, and $n_{fp}=4$. These parameters were chosen to ensure good divertor characteristics following coil optimization.

\subsection{Weighted Squared Flux \label{sec:Method-WSF}}

\begin{figure}
    \centering
    \includegraphics[width=0.5\textwidth]{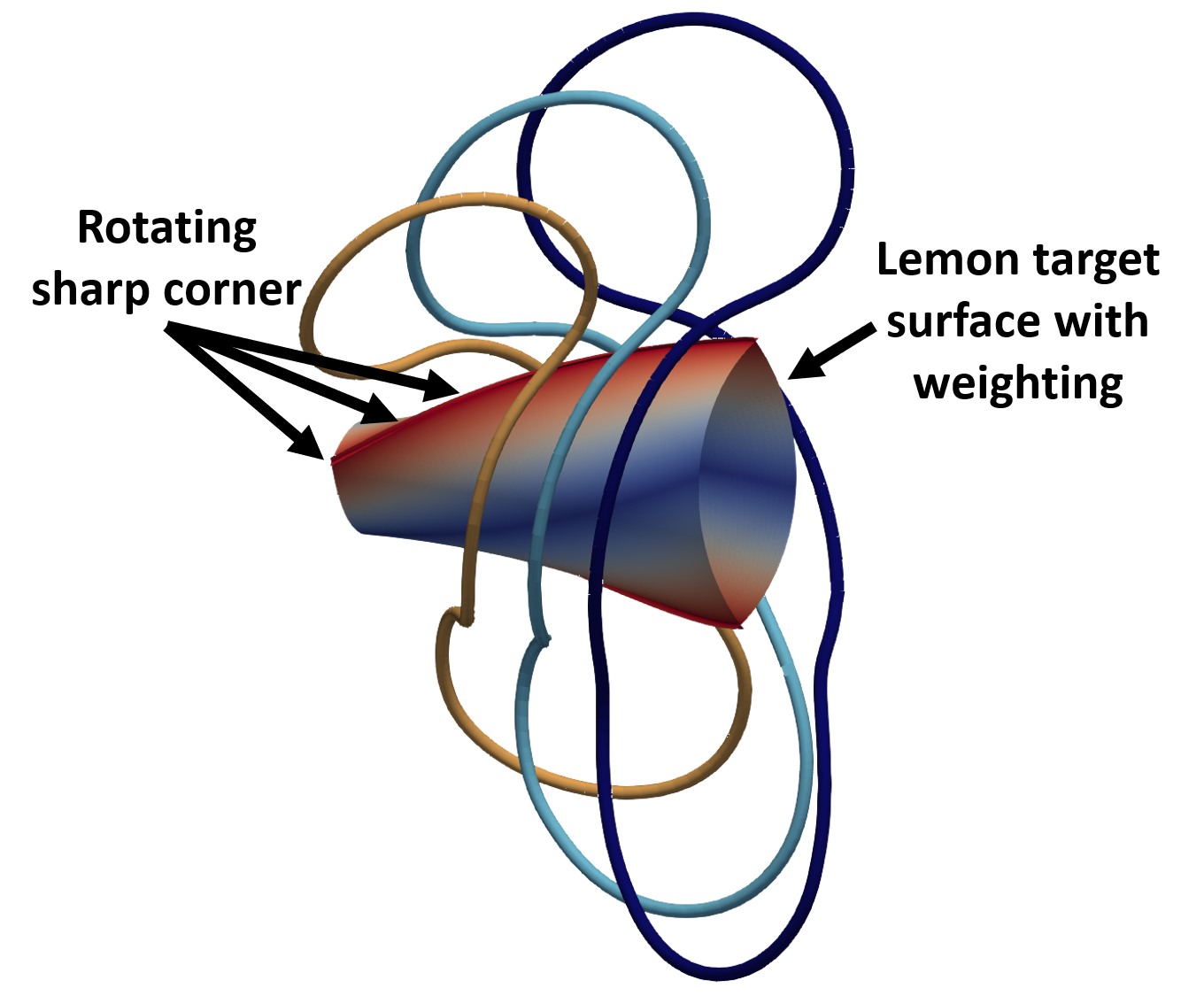}
    \caption{Lemon target surface showing the weight used for the weighted squared flux objective and an optimized coil set. One half field period is shown. Near the sharp corners of the lemon, the weight approaches 1 (red). Away from the corners, the weight decreases according to eqn. \ref{eqn:WSF_weights} (blue). The coils shown are the result of weighted squared flux optimization and correspond to the Poincare section in figure \ref{fig:all_poincares}(d).}
    \label{fig:3d_wsf_coils}
\end{figure}

Typical $B_n$ minimization was largely successful, though large $B_n$ errors persisted near the sharp corners of the lemon. To remedy this, a weighted squared flux (WSF) was introduced to emphasize field accuracy near the sharp corners of the lemon, de-prioritizing field accuracy at smoother parts of the target surface.

In the weighted approach, the quadratic flux integral was weighted inversely with distance from the corners of the lemon:
\begin{equation}
f_w = \int w\left(\mathbf{x}\right) B_n\left( \mathbf{x} \right) ^2 \, dA, 
\label{eqn:WSF}
\end{equation}
with
\begin{equation}
w\left(\mathbf{x}\right) = \left( 1-\frac{|\mathbf{x}-\mathbf{x}_{0}|}{d_{max}} \right)^{p}    ,
\label{eqn:WSF_weights}
\end{equation}
where $\mathbf{x}$ is a location on the lemon surface, $\mathbf{x}_{0}$ is location of the closest sharp corner, $d_{max}$ is the maximum surface-intersection distance, and $p$ is a coefficient controlling the locality of the weighting. As $p\to 0$ we revert to typical squared flux, while as $p \to \infty$ only points near the intersection are targeted. 

An example of the weights on the surface is illustrated in Fig.~\ref{fig:3d_wsf_coils}.

\subsection{Manifold Optimization \label{sec:Method-Maniopt}}

\begin{figure}
    \centering
    \includegraphics[width=0.5\textwidth]{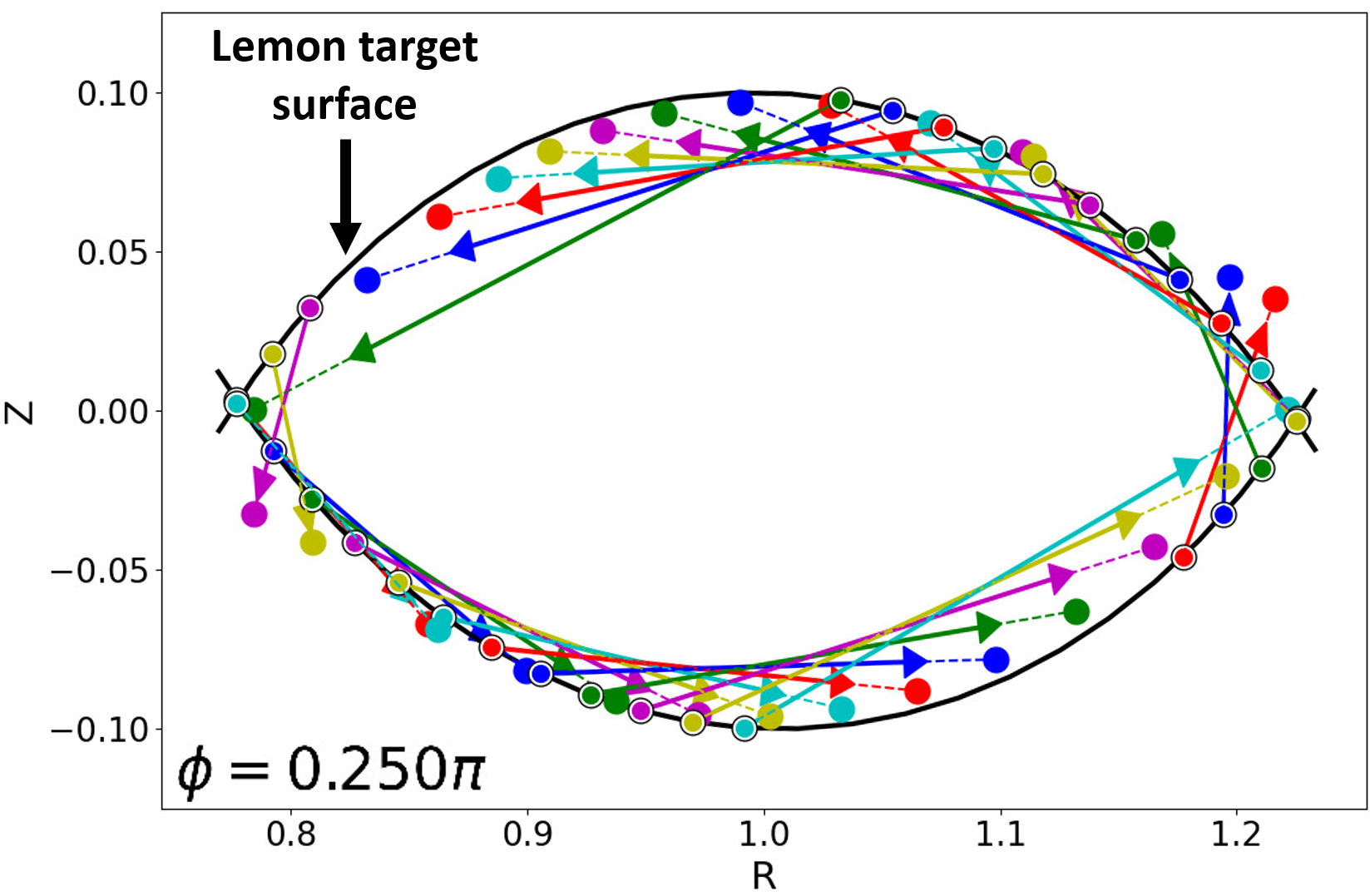}
    \caption{Illustration of manifold optimization}
    \label{fig:maniopt_method}
\end{figure}

To further enhance the separatrix quality of the lemon, a novel optimization technique, termed manifold optimization, is employed. This method penalizes the deviation of field lines originating on a target surface from that surface after tracing through one full field period. This method is intended to reduce chaos in the diverted field line region.

Manifold optimization is illustrated in Fig. \ref{fig:maniopt_method}. To begin, we use an array of points on the target surface for a single $\phi$ angle, here chosen to be $\phi_0=0.25\pi$. Magnetic field lines started from these points are then traced for one field period, with integration ending at $\phi_{end}=\phi_0+(2\pi/n_{fp})$. The endpoints of these field lines are then used to compute the deviation of the field lines at $\phi_{end}$ from the target surface.
The objective function to be minimized is
\begin{equation}
    J_{\text{maniopt}} = \sum_i \min|\mathbf{x}_i-\mathbf{S}|
    \label{eqn:Maniopt_penalty_function}
\end{equation}
where $\mathbf{x}_i$ is the endpoint coordinate of the $i$th traced field line and $\mathbf{S}$ is the closest point on the target surface. 

In this work, manifold optimization was used to improve the results of previously-optimized coils. Indeed, manifold optimization does not naively work with cold-start optimization when the coils are initially planar: in a purely toroidal field, all field lines start back at their initial location after one field period, rendering \ref{eqn:Maniopt_penalty_function} equal to 0. 

An argument for why manifold optimization may be more effective for chaos suppression than $B_n$ miminization is that the latter treats resonant field errors as no different from nonresonant errors. Resonant errors typically result in large changes to magnetic field line trajectories (for example, opening up magnetic islands) in response to small error fields. By directly targeting the properties of field line trajectories,  manifold optimization may more efficiently eliminate the resonant field errors that stochastize the edge magnetic field.


A curious observation is that the manifold optimization objective only depends on the field line locations at a single $\phi$ angle (and its field-period-symmetric location $\phi_{end} = \phi_0 + 2\pi/n_{fp}$.) In effect, the magnetic field lines may move freely between $\phi_0$ and $\phi_{end}$, providing equilibrium flexibility. Manifold optimization could perhaps be modified to also control the field line locations at intermediate angles.
This possibility is not explored further in this paper, though is an interesting topic for future work.

This method was developed in SIMSOPT with PyOculus, a Python package for magnetic field topology analysis \cite{pyoculus}.

\subsection{Fixed-point analysis \label{sec:Method-FPs}}

When coils are optimized using the three methods above, some amount of chaos will remain around the separatrix. To assess the amount of this chaos we use a measure based on periodic field lines, known as fixed points of the Poincare map.
These fixed points can be classified as X-points, where neighboring field lines converge or diverge in a hyperbolic structure (indicative of separatrix boundaries), or O-points, about which field lines rotate in elliptic orbits (often associated with magnetic islands). These points can be used to understand the characteristics of a stellarator divertor.
    
Greene's residue quantifies the stability of these fixed points by analyzing the behavior of field lines in a neighborhood of a periodic field line \cite{Greene1979}. 
A residue between 0 and 1 indicates an elliptic (stable) point, while values $<0$ or $>1$ imply a hyperbolic (unstable) point, potentially leading to chaotic regions. This metric has been used to optimize stellarator fields by targeting and minimizing island widths, thereby reducing chaos \cite{hanson1984elimination}.

Fixed point analysis is used here to assess the degree of chaos achieved with each optimization method. Following coil optimization, X and O-points near the sharp corner of the lemon were identified, and Greene's residue of each resonance computed to quantify local stability. This provides a metric of the degree of chaos in the resulting magnetic fields.

Chaos is the result of island (or resonance) overlap. Each rotating corner of the lemon requires two field periods to complete one full poloidal rotation, corresponding to an $m=2$ resonance, where $m$ is the number of field period mappings necessary for a point to return to itself. Consequently, the resonances used for chaos quantification are those closest to the $m = 2$ resonance. Fixed points with $m=2$, 3, and 4 are found, and their Greene's residue computed, to quantify the degree of chaos in each optimized coil configuration.

Greene's residue of fixed points was also explored as an optimization target to mitigate chaos. Although not detailed here, it proved to be overly sensitive to coil perturbations for effectively reducing chaos in the diverted field line region, despite performing well near the core.

Another metric considered for chaos quantification was turnstile area \cite{Turnstiles}, though this, too was found to be too sensitive to coil perturbations.

Here, we describe the sensitivity of the system to perturbations in the coils. Locating the $m=2$ resonances in the Lemon configuration is highly sensitive to the initial guess, due to the rapid divergence of nearby field lines. This extreme sensitivity is evidenced by Greene's residue values on the order of $10^6$ (see Table \ref{table:FP_resonances}), which indicate exponential divergence from the fixed point \cite{Davies_2025}. During optimizations, any changes to the coils must be extremely small to ensure that the fixed point in the new field remains within the convergence region of the rotator method. Since fixed-point finding in PyOculus \cite{pyoculus} relies on a Newton method—which has a limited region of convergence—optimizing the properties of the main period-2 resonance becomes computationally infeasible.

\section{Results \label{sec:Results}}

\begin{figure}
    \centering
    \includegraphics[width=0.5\textwidth]{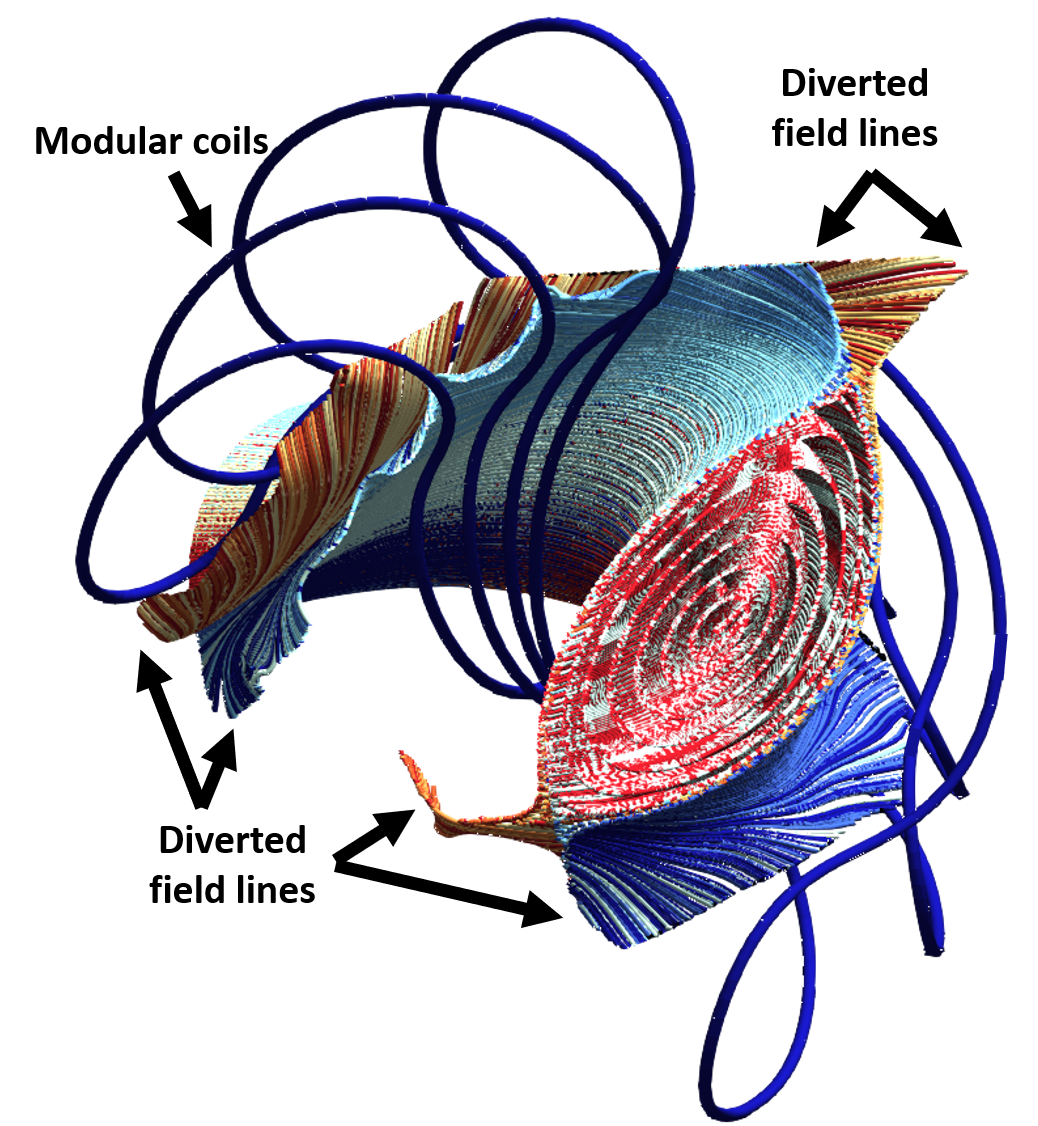}
    \caption{Three-dimensional rendering of the rotating lemon divertor structure. The divertor legs are well-separated and non-chaotic.
    }
    \label{fig:3d_lemon_coils}
\end{figure}

\begin{figure*}
\centering
\includegraphics[width=1.00\textwidth]{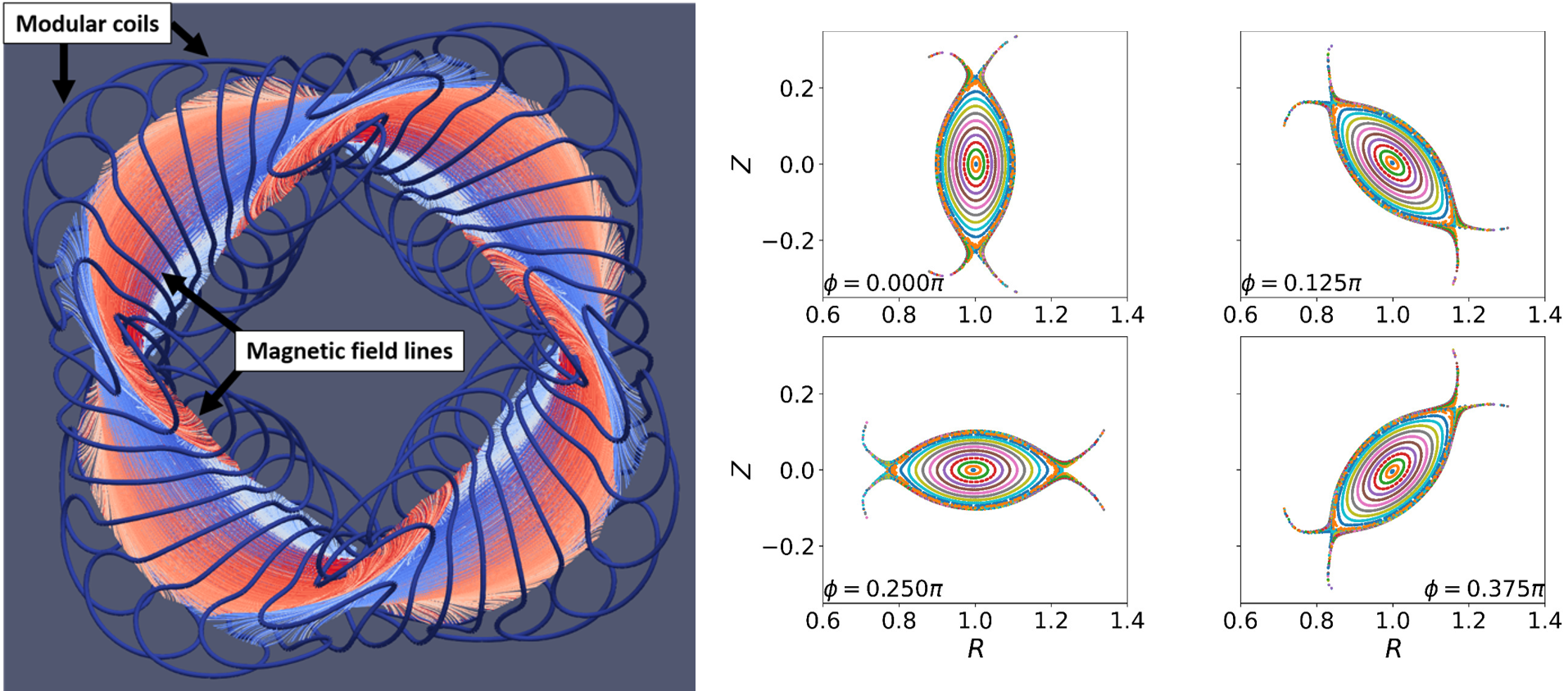}
\caption{(a): Coil set of the rotating lemon after typical $B_n$ minimization with field lines shown. (b): Poincare sections for the lemon coil set at several toroidal angles.}
\label{fig:coils+poincares}
\end{figure*}

\begin{figure*}
\centering
\includegraphics[width=1.00\textwidth]{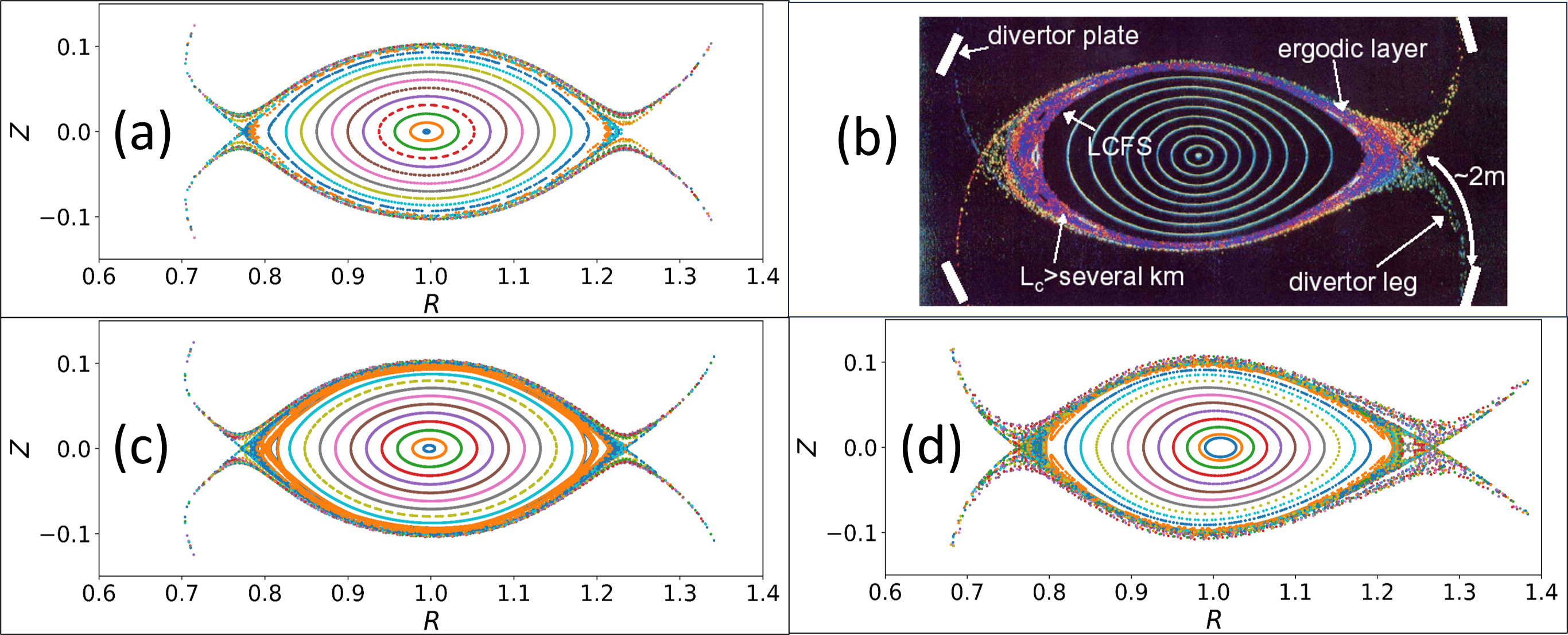}
\caption{\Poincare sections of (a) the lemon, (c) Manifold-optimized, (d) Weighted Squared Flux coil sets and (b) the LHD divertor for comparison. 
Panel (b) is reproduced from Ref \onlinecite{LHD_figure}.
}
\label{fig:all_poincares}
\end{figure*}

The result of straightforward $B_n$ minimization on the rotating lemon may be seen in Figs. \ref{fig:3d_lemon_coils}-\ref{fig:coils+poincares}. The coil set pictured here, hereafter referred to as the lemon coil set, is used as our baseline result for comparison with other methods.
It includes four unique coil shapes per half period, for 32 coils in total.

The diverted field line properties of the lemon coil set are reminiscent of a standard tokamak X-point divertor. 
As can be seen in figures \ref{fig:3d_lemon_coils} and Fig. \ref{fig:coils+poincares}, the divertor legs fan out crisply from the sharp edge of the lemon.
That is, field lines just outside the lemon boundary flow to the X-lines and turn a corner, then move far from the plasma. 
Just as in LHD, baffles could be placed close to the divertor legs, with a thin gap around the X-lines for outflow from the plasma.
This closed geometry would prevent neutrals and impurities originating from the divertor target plates from having a line of sight to the core, concentrating neutrals for efficient pumping.
When compared with the LHD divertor, reproduced here in Fig. \ref{fig:all_poincares}(b), the lemon's diverted field line region (Fig. \ref{fig:coils+poincares}(b)) is considerably less chaotic. 

The lemon coil set has some distinctive differences compared to typical modular coils. At its widest point, each coil has a section which runs nearly parallel to the sharp edge of the lemon. This appears to be an emulation of helical coils by the modular coils. 

A coil set generated using the weighted squared flux method, hereafter referred to as the WSF coil set, may be seen in Fig. \ref{fig:3d_wsf_coils}, with the Poincare section appearing in Fig. \ref{fig:all_poincares}d. The WSF coil set was optimized using randomized sampling over weights and thresholds of eqns. \ref{eqn:Coil_eng_penalty_function} and \ref{eqn:Maniopt_penalty_function}. The WSF coil set features improved coil engineering metrics, as shown in Table. \ref{table:CO_metrics} at the expense of higher separatrix chaos. Indeed, the diverted field line region of the WSF coil set is similar to that of LHD, Fig. \ref{fig:all_poincares}b. Further, the lemon target shape is not as closely matched for the WSF coil set, though this is not surprising as the WSF method emphasizes field correctness near the sharp corners. Overall, the WSF coil set produces an LHD-like divertor with more engineering-friendly modular coils.

A coil set generated using manifold optimization, hereafter referred to as the MO coil set, is not shown here as it is visually indistinguishable from the original lemon coils, with differences at the mm level. The MO coil set was produced by applying manifold optimization with the lemon coil set as an initial condition. A \Poincare section of the MO coil set may be seen in Fig. \ref{fig:all_poincares}c. The MO coil set exhibits reduced chaos in the separatrix according to the residues in Table \ref{table:FP_resonances}, though visually the \Poincare plot is similar.

Coil complexity metrics are compared in Table \ref{table:CO_metrics}. The initial lemon coil set, Fig. \ref{fig:3d_lemon_coils}, has higher curvature and lower coil-coil clearances. The WSF method, in contrast, relaxes each considered coil engineering penalty by a factor of two and uses only three coils per half field period. The MO coil set is virtually identical to the lemon coil set.

\begin{figure}
    \centering
    \includegraphics[width=0.5\textwidth]{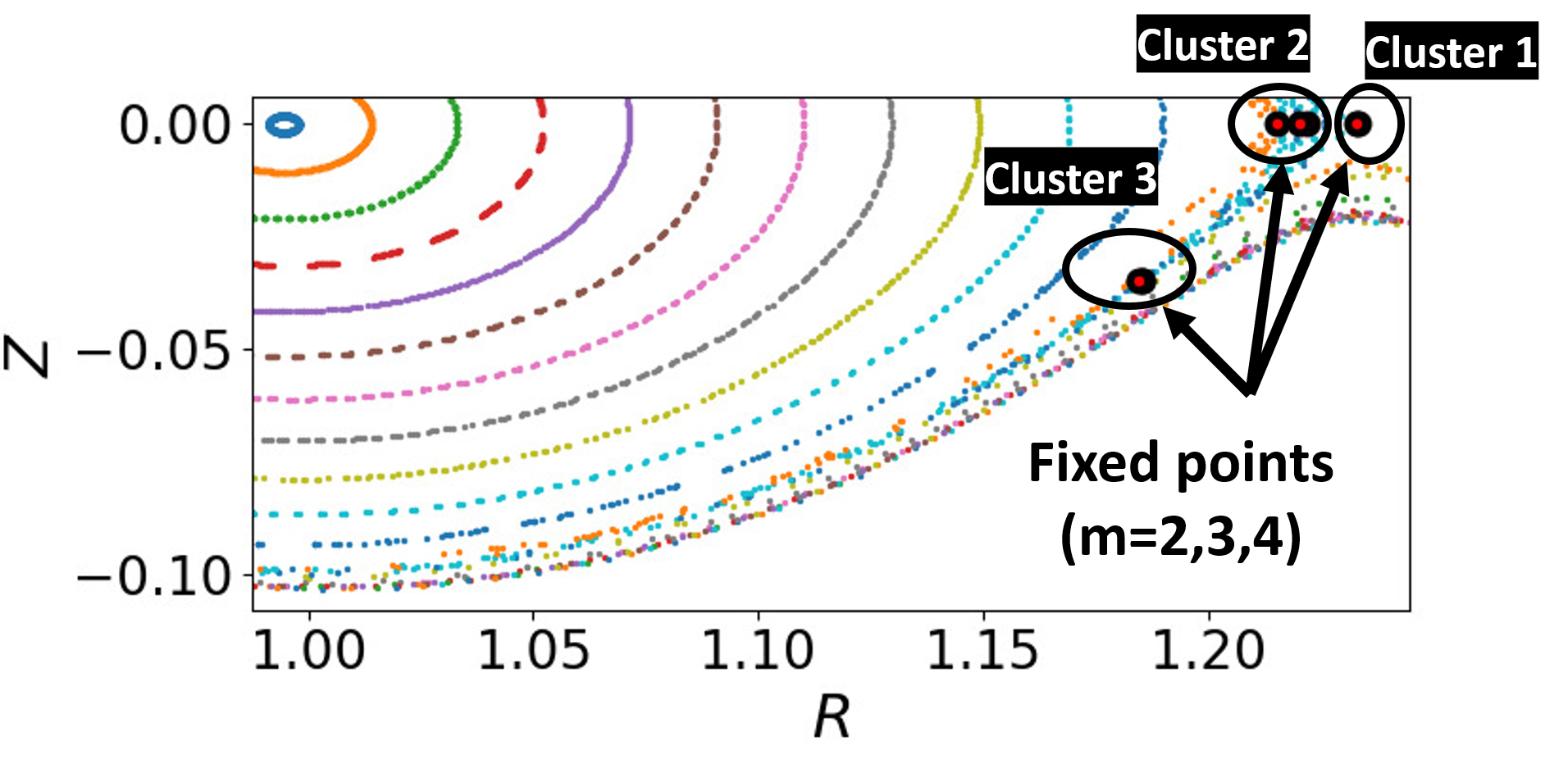}
    \caption{Locations of selected fixed points for quantifying divertor chaos, grouped into clusters by spatial proximity. Cluster~1 (primary resonance branch) includes $m=2$ and $m=3$ fixed points at $R \in (1.23295, 1.2330)$ and $Z = 0$. Cluster~2 comprises $m=2$ and $4$ fixed points at $R \in (1.215, 1.225)$ and $Z = 0$. Cluster~3 contains $m=3$ and $4$ periodic fixed points at $R \in (1.1845, 1.1855)$ and $Z \in (-0.0344, -0.0350)$. Positions vary slightly across coil configurations. Greene's residues, assessing stability and chaotic behavior, appear in Table~\ref{table:FP_resonances}.}
    \label{fig:Lemon_FPs}
\end{figure}

Greene's residues for fixed points in the diverted field line region are shown in Table \ref{table:FP_resonances}, and their locations are shown in Fig. \ref{fig:Lemon_FPs}. The $m=2$ fixed point is the main resonance of the lemon's divertor concept as the corners rotate with $m=2$ periodicity. There is an additional $m=3$ fixed point which lies within 10 $\mu$m of the $m=2$ fixed point. These two fixed points, hereafter referred to as the main fixed points, are inextricably linked to the chaos of the diverted field lines. On the other hand, the fixed points of cluster 2 and cluster 3 from Fig. \ref{fig:Lemon_FPs} are linked to chaos both in the diverted field line region and near the last closed flux surface. Fixed points in clusters 2 and 3 are referred to as the secondary fixed points.

For the lemon coil set, the primary fixed points exhibit large Greene’s residues, while secondary fixed points show small values. Consequently, field lines in the divertor region transit rapidly. Despite this, the separatrix remains largely free of chaos.

For the MO coil set, residues of the primary fixed points are reduced by several orders of magnitude compared to the lemon case (Table~\ref{table:FP_resonances}), resulting in a less chaotic divertor region. However, the secondary fixed points have larger residues, increasing stochasticity near the last closed flux surface.

The simple fixed-point analysis used in Table \ref{table:FP_resonances} fails to fully capture the chaos of the WSF coil set, as multiple overlapping resonances contribute significantly to the stochastic layer.

\begin{table}[htbp]
\centering
\caption{Greene's residues of resonant fixed-point clusters near the divertor region (see Fig.~\ref{fig:Lemon_FPs}). Only lemon and MO configurations are shown to illustrate the chaos suppression achieved through manifold optimization.}
\begin{tabular}{
  l
  l
  r
  r
  r
  r
}
\toprule
\textbf{Coils} & 
\textbf{Cluster} & 
{$m=2$} & 
{$3$} & 
{$4$} \\
\midrule
Lemon & \qquad 1 & -3.5E+06 &   2.7E+04 &\\
MO    &   & -5.9E+00 &   2.1E+00 &\\
\midrule
Lemon & \qquad 2 &  2.9E+00 &  &  -6.4E+02\\
MO    &   &  2.1E+00 &  &  -3.3E+00\\
\midrule
Lemon & \qquad 3 &   & 2.7E-01 &  -4.3E+00 \\
MO    &   &   &  -3.0E+04 &  2.3E+02\\
\bottomrule
\end{tabular}
\label{table:FP_resonances}
\end{table}

\begin{table*}[htbp]
\centering
\caption{Engineering performance comparison of the coil sets across four metrics: the number of coils per half field period, total coil set length (m), minimum coil-coil distance (m), maximum coil curvature (m$^{-1}$), and maximum coil mean squared curvature (m$^{-2}$)}
\begin{tabular}{
  lccccc
}
\toprule
\textbf{Coil Set} & 
{\textbf{$n_{coils}$}} & 
{\textbf{Length (m)}} & 
{\textbf{Min. cc dist. (m)}} & 
{\textbf{Max. curv. (m$^{-1}$)}} & 
{\textbf{Max. mean sq. curv. (m$^{-2}$)}} \\
\midrule
Lemon  & 4 & 123.57 & 0.046 & 9.87 & 22.36 \\
WSF    & 4 & 82.35  & 0.11 & 4.80  & 12.22 \\
MO     & 3 & 123.57 & 0.046 & 9.87 & 22.36 \\
\bottomrule
\end{tabular}
\label{table:CO_metrics}
\end{table*}

\section{Discussion}
\label{sec:discussion}

In this work we have presented a new option for a stellarator divertor: a helical divertor produced with optimized modular coils.
This solution combines the principal advantage of the helical divertor -- the possibility of tight baffling to confine neutrals in front of the pumps -- with the advantages of modular coils: the ability to wind coils in a factory off-site and transport them to the final location rather than wind them on the vacuum chamber.
We also demonstrated that the high level of chaos around the separatrix of LHD's divertor is not necessary in a helical divertor -- in fact helical divertors can be found with much reduced chaos.
The key to the method was performing standard modular coil optimization using a target plasma surface with a toroidally continuous sharp edge.

In this paper we also presented manifold optimization, a new objective function which can be incorporated in coil design.
Manifold optimization may have the potential to produce separatrices with low chaos, although in the example here the resulting configuration was very sensitive to coil perturbations.  Nonetheless, it represents a new option for reducing stochasticity and is computational efficient. 

The manifold-optimized coil set's sensitivity to perturbations underscores a fundamental challenge in stellarator divertor design. Because diverted field lines must pass close to the diverting coils, their separatrices become inherently fragile. It remains to be shown that a divertor robust to coil perturbations can be realized. In the computational design of stellarator divertors, this fragility frequently drives field-line tracing to the limits of numerical precision.

Additional work is required to develop this divertor concept.
Perhaps most significantly, the target plasma surface in this work was not optimized for good confinement or stability.
Such optimization of the plasma boundary has previously been performed using equilibrium codes with a Fourier description of the boundary, resulting in a smooth plasma surface.
Therefore a key question is how to optimize plasma boundary shapes with both good core physics properties and sharp edges.
Ref \onlinecite{gaur2025omnigenous} presented an approach to this question,  but the edge field behavior was significantly different from the sharp separatrix shown here, perhaps because the boundaries in Ref. \onlinecite{gaur2025omnigenous} did not have a truly discontinuous normal.
A related question is whether modern quasisymmetric (QS) or quasi-isodynamic (QI) plasma geometries are compatible with the helical divertor here.
Typically QS and QI plasma boundaries have a high-curvature ridge which extends toroidally for about one field period, but not around the entire torus.
Thus it remains to be seen whether a helical divertor like the one here can be generated for a QS or QI geometry.

Further development of this divertor concept will also require more sophisticated modeling to study many topics.
These include the width of the heat flux layer, behavior of neutrals in the presence with realistic wall and baffle geometry, sensitivity to plasma beta, and detachment access.

Another question for future work is the optimal level of chaos in the separatrix.
The chaos in LHD's separatrix has been suggested as a reason it has been hard to achieve stable detachment \cite{feng2011comparison}, suggesting that reduced chaos would be beneficial.
On the other hand, chaos in the separatrix likely increases the width of the heat flux layer, reducing the peak heat flux on the divertor plates.
Research is required to determine the optimal balance between these considerations.





\section*{Acknowledgments}
T. E. was supported by the U.S. Department of Energy Fusion Energy Sciences Postdoctoral Research Program
administered by the Oak Ridge Institute for Science and Education (ORISE) for the DOE. ORISE is managed by Oak Ridge
Associated Universities (ORAU) under DOE contract number DE-SC0014664. 
This work was also supported by the U.S. Department of Energy, Office of Science, Office of Fusion Energy Sciences under Award DE-FG02-93ER54197.

\section*{Data availability statement}
Data available on request from the authors.

\bibliography{references}

\end{document}